\documentclass[12pt,letterpaper]{article}          %% LaTeX 2e (preferred)
\usepackage{osajnl}
\usepackage[draft]{hyperref} % optional

\def \apj {The Astrophysical Journal } %{ApJ}
\def \apjs {The Astrophysical Journal Supplements } %{ApJS}
\newcommand{\citeNP}[1]{\cite{#1}}

\begin{document}

\title{Polarimetric Calibration of Large-Aperture Telescopes I: The
  Beam-Expansion Method}

\author{Hector Socas-Navarro}
%% for REVTeX4, each author name can be set in a separate \author{} field

\address{3450 Mitchell Ln, Boulder CO, 80307-3000}

\begin{abstract}This paper describes a concept for the high-accuracy absolute
calibration of the instrumental polarization introduced by the primary mirror
of a large-aperture telescope. This procedure requires a small aperture with
polarization calibration optics (e.g., mounted on the dome) followed by a
lens that opens the beam to illuminate the entire surface of the mirror. The
Jones matrix corresponding to this calibration setup (with a
diverging incident beam) is related to that of the normal observing setup
(with a collimated incident beam) by an approximate correction
term. Numerical models of parabolic on-axis and off-axis mirrors with
surface imperfections are used to explore its accuracy.
\end{abstract}

\ocis{120.2130, 120.5410, 120.4640, 350.1260}
%\ocis{000.0000, 999.9999.}% REPLACE WITH CORRECT OCIS CODES FOR YOUR ARTICLE
                          % NOTE: \ocis{} IS ALIASED TO \pacs{} BUT MUST
                          % FORMAT THE TERMS CORRECTLY FOR EACH JOURNAL

\maketitle %% NULL FUNCTION WITH LATEX 2e; required for REVTeX4

\section{Introduction}
\label{sec:intro}

Astronomical polarimetry requires frequent calibration operations to remove
the instrumental polarization introduced by the various optical elements
encountered by the light beam along its path. This contamination is usually
determined by 
placing calibration optics early in the light path, which is used to feed
light in a known state of polarization into the instrument. By measuring the
polarization of the light that comes out of the system, it is possible
to characterize it in terms of its Jones matrix (or, alternatively, the
Mueller matrix if one is using the Stokes formalism). 

Obviously, we can only characterize and calibrate the optical elements that
the beam encounters {\it after} the polarization calibration
optics. Therefore, calibrating a telescope requires placing such optics at
the telescope entrance, before the first reflection on the primary mirror
(M1) occurs. This approach\cite{SLMP+97} is being successfully employed
for the Dunn Solar 
Telescope (at the Sacramento Peak Observatory, managed by the National Solar
Observatory) and the German VTT on the island of Tenerife (at the
Observatorio del Teide of the Instituto de Astrof\' \i sica de Canarias). In
both cases, an array of linear polarizers and retarders are slided in the
light path, on top of the entrance window, for calibration
operations. Separate calibrations are obtained for the telescope 
and the instrument, so that the former does not need to be done as
frequently. The Jones matrix of the complete system is then obtained 
as the product of the telescope and instrument matrices. 

Unfortunately, entrance window polarizers are not practical for apertures
larger than $\sim$1~m diameter. In the past this has not been a major concern
because: a)solar telescopes have apertures that do not exceed 1~m; and
b)large astronomical telescopes have not been used for polarization
measurements. However, this scenario is starting to change. Polarimetry is
proving to be a very powerful tool to explore a broad range of Astrophysical
problems\cite{WS02}, resulting in
a rapidly increasing interest to 
develop polarimeters for existing large telescopes. Second, the development
of the Advanced Technology Solar Telescope (ATST)\cite{KRH+02,KRK+03}
demands a reliable method to calibrate a large telescope for
high accuracy spectro-polarimetry.

Solar telescopes have been calibrated in the past by observing magnetic
structures and making assumptions on the underlying physics. This poses
important challenges, however, especially when pushing the envolope towards
new observational domains. Consider for example the ATST, which is intended
to do polarimetry at the $\sim$10$^{-4}$ accuracy level. One of the common
assumptions that is usually made in solar polarimetry is to consider that the
continuum radiation is unpolarized. However, scattering processes can
polarize the continuum and generate signals of the order of $\sim$1\% in the
blue side of the visible spectrum. It has been stated \cite{G03} that ``the
direct observation of the polarisation of the continuous radiation still is a
major outstanding observational challenge''. Considerations on the symmetry
properties of Stokes profiles are not appropriate either. Gradients in the
line-of-sight velocity or magnetic field introduce spurious asymmetries that
can invalidate these assumptions. Furthermore, some physical processes
operating in the atoms are known to induce Stokes asymmetries even in the
absence of gradients. Spectral lines forming in the incomplete Paschen-Back
regime become asymmetric\cite{SNTBLdI04}. Moreover, the
alignment-to-orientation conversion mechanism\cite{KMN84} may also cause
profile asymmetries. 

In summary, it is very important to have an absolute calibration that does
not rely on preconceived ideas on the objects under study. This is specially
true when the instrumentation is being used to explore new scientific
realms. The present work has been motivated by the 
challenge of calibrating the 4-meter primary mirror of the ATST to meet its
very stringet polarization requirements. It might also be possible to use the
calibration method proposed here in other existing large-aperture
telescopes. However, the actual design of a practical implementation is beyond
the scope of this paper. The main point of this work is to show that a
calibration setup with inclined beam incidence can be used to measure the
polarization properties of a large-aperture primary mirror. Geometrical effects
can be calculated and removed from the measured Jones or Mueller matrices
resulting in a good approximation to such matrices in normal observing
conditions. 

\section{The calibration setup}

In this paper we shall consider two different configurations: the normal
observing setup (OS) and the calibration setup (CS) proposed
here. Fig~\ref{fig:setups} shows a 
schematic representation of the OS and CS for an on-axis M1 mirror. Our
ultimate goal is to determine the Jones matrix of M1 in the OS ($M^{OS}$). A
direct measurement of $M^{OS}$ would require polarization optics of the same
diameter as the telescope aperture, which is not practical due to technical
difficulties. The $M^{CS}$ matrix, on the other hand, can be determined by
mounting appropriate calibration optics (and a mechanical control system)
at a height $H_{cal}$ over M1 (as shown in Fig~\ref{fig:setups}, right). A
small aperture on the dome is probably an ideal location for it.

The CS requires some (small) amount of additional optics with respect to the
OS. At least two lenses are required: one at the entrance to open the beam
and another at the detector for imaging. These additional elements should be
designed with stringent polarization requirements. Aberrations, chromatism
and other image imperfections can be tolerated in these components, which
will only 
be used for calibration. Even if they introduce some residual polarization,
this 
should be easily measurable in the laboratory and removed from the M1 Jones
matrix. 

The size of the calibration optics affects the accuracy of the
calibration. In the limit where it fills the entire telescope
aperture, the CS and 
the OS are identical and no correction is needed. As the calibration aperture
becomes smaller, the incidence angles increase resulting in larger
corrections (and errors). The simulations presented in this paper consider
the pessimistic limit in which the calibration optics has a diameter
approaching zero (a pinhole aperture).

\section{Basic relations}

For the calculations
in this paper it is convenient to use cylindrical coordinates with the
vertical axis along the propagation direction of the incident light beam. The
radial coordinate $\rho$ is the distance from the center of the aperture and
$\phi$ is the azimuth angle measured from an arbitrary reference. Any given
point on the surface of the M1 mirror at coordinates ($\rho$, $\phi$) is 
characterized by its complex refraction index $N(\rho,\phi)$. We shall
consider here the behavior of a monochromatic plane wave. This will allow us
to describe the instrumental polarization of M1 in terms of its Jones matrix
$M(\rho,\phi)$. Appendix~\ref{sec:appendix} gives the Mueller equivalent 
of the most important Jones matrices derived in this work.

Locally, the behavior of M1 in the vicinity of ($\rho$, $\phi$) may be
approximated by a reflection on a flat mirror of homogeneous refraction index
$N=n+i\kappa$. This process adopts a very simple form in the reference system
of the plane of incidence (the plane formed by the incoming ray and the
surface normal, see Fig~\ref{fig:angles}). Let us denote the components 
in the plane of incidence with the subindex $p$ and those perpendicular to it
with $s$. In this frame, the Jones matrix of the reflection, ${\cal M}_{ps}$,
is simply\citeNP{BW75}: 

\begin{equation}
{\cal M}_{ps}=\left ( \begin{array}{ccc}
                  r_p & ~~ & 0 \\
                   0  &     & r_s \end{array} \right ) \, .
\end{equation}
Assuming that the refraction index of air is 1, $r_p$ and $r_s$ are
given by:

\begin{equation}
\label{rp}
r_p={  {\sqrt{N^2 - \sin^2 \theta} - N^2 \cos \theta } \over
     {\sqrt{N^2 - \sin^2 \theta} + N^2 \cos \theta } } \,
\end{equation}

\begin{equation}
r_s={\cos \theta - {\sqrt{N^2 - \sin^2 \theta} } \over
     {\cos \theta + \sqrt{N^2 - \sin^2 \theta} } } \, ,
\end{equation}
where $\theta$ is the angle of incidence. This matrix can be transformed to the
global reference frame ($\rho$, $\phi$):

\begin{equation}
\label{mirror1}
M(\rho,\phi) = R(-\phi) {\cal M}_{ps} R(\phi) \, ,
\end{equation}
where $R(\phi)$ is the usual rotation matrix. Writing down $M(\rho,\phi)$
explicitly: 
\begin{equation}
\label{explicit}
M(\rho,\phi) = \left ( \begin{array}{ccc} 
    r_p \cos^2 \phi + r_s \sin^2 \phi & ~~~~~~ & (r_p+r_s) \sin \phi \cos \phi \\
    (r_p+r_s) \sin \phi \cos \phi &  & r_p \sin^2 \phi + r_s \cos^2 \phi
    \end{array} \right )
\end{equation}

\section{Mirror models}

\subsection{On-axis mirror}
\label{sec:onaxis}

Consider an axi-symmetric mirror illuminated by a collimated beam (OS). The
angle of incidence $\theta$ is constant along concentric rings in the
mirror ($\theta=\theta(\rho)$). For example, a parabolic mirror of focal
length $F$ is characterized by the condition:

\begin{equation}
\tan \theta = {\rho \over 2 F} \, .
\end{equation}

Suppose that the complex refraction index is constant over the surface of the
mirror (perfect mirror). In this case, $r_p$ and $r_s$ are only functions of
$\rho$ (because 
$\theta=\theta(\rho)$). The dependences of $M(\rho,\phi)$ in
Eq~(\ref{explicit}) are easily separable. The Jones matrix of a thin ring of
radius $\rho$ is simply:

\begin{equation}
\label{ring}
M(\rho)= \rho \int_0^{2 \pi} M(\rho,\phi) d \phi =
\rho \pi \left ( \begin{array}{ccc} (rs+rp) & ~~ & 0 \\
                                           0   & ~~ & (rs+rp) \end{array}
\right ) \, .
\end{equation}

Eq~(\ref{ring}) represents the Jones matrix of a non-polarizing system
(identical reflectivity and retardance for both components of the electric
field). This is a well-known property of axi-symmetric systems. Notice,
however, that the symmetry is broken if one observes away from the center of
the field of view. Imperfections in the mirror (irregularities in the
refraction index caused by coating degradation, dust, etc) may also
break the symmetry of the system and introduce instrumental polarization.

Let us now turn to the more general case of an imperfect mirror, defined as
one with $N=N(\rho,\phi)$. If this is the case then $r_p$ and $r_s$ vary
across the mirror and Eq~(\ref{ring}) is no longer valid. We seek to
determine a suitable calibration by means of the (measurable) $M^{CS}$
matrix. The differences between $M^{OS}$ and $M^{CS}$ are due to the
different incidence angle of the beam (represented in
Fig~\ref{fig:angles}). We can expand $r_p$ in a power series of
$\alpha$ as:

\begin{equation}
\label{expandrp}
r_p(\theta_{CS})=r_p(\theta_{OS}-\alpha) = r_p(\theta_{OS}) -
\alpha  {d r_p \over d \theta } \vert _{\theta_{OS}}
+{\alpha^2 \over 2 } {d^2 r_p \over d \theta^2 } \vert _{\theta_{OS}}
+ \dots \, ,
\end{equation}
and similarly for $r_s$. Inserting this expansion into
Eq~(\ref{explicit}) we obtain:

\begin{eqnarray}
\label{expandM}
M^{OS}(\rho,\phi) \simeq M^{CS}(\rho,\phi) - \alpha \left ( \begin{array}{ccc}
  d_{1,p} \cos^2 \phi + d_{1,s} \sin^2 \phi & ~~~~ &  (d_{1,p}+d_{1,s}) \sin \phi \cos \phi \\
  (d_{1,p}+d_{1,s}) \sin \phi \cos \phi    &    & d_{1,p} \sin^2 \phi + d_{1,s} \cos^2 \phi
				\end{array} \right ) \nonumber \\
+ {\alpha^2 \over 2} \left ( \begin{array}{ccc}
  d_{2,p} \cos^2 \phi + d_{2,s} \sin^2 \phi & ~~~~ &  (d_{2,p}+d_{2,s}) \sin \phi \cos \phi \\
  (d_{2,p}+d_{2,s}) \sin \phi \cos \phi    &    & d_{2,p} \sin^2 \phi + d_{2,s} \cos^2 \phi
				\end{array} \right ) + \dots
 \, ,
\end{eqnarray}
where $d_{i,p}$ and $d_{i,s}$ have been introduced for notational simplicity:
\begin{equation}
\label{dp}
d_{i,p}={d^i r_p \over d \theta^i } \vert _{\theta_{OS}} \, ,
\end{equation}
(and similarly for $d_{i,s}$). In the equations above, $\alpha$, $d_{i,p}$ and $d_{i,s}$
 are all functions of $(\rho,\phi)$. The angle $\alpha$ can be easily
 determined from geometrical considerations. However, $d_{i,p}$ and $d_{i,s}$ are
 affected by imperfections in the mirror that change over time. Let us
 separate $d_{i,p}$ into two components: a nominal $\hat d_{i,p}$ derived from
 Eqs~(\ref{rp}) and~(\ref{dp}) with a theoretical refraction index  $N^{nom}$
 (e.g., from manufacturer specifications), and an unknown $\delta d_{i,p}$ due to
 coating degradation, dust accumulation, etc:
\begin{equation}
\label{splitd}
d_{i,p}=\hat d_{i,p} + \delta d_{i,p} \, ,
\end{equation}
(and similarly for $d_{i,s}$). Inserting this into Eq~(\ref{expandM}) and
integrating over the entire mirror surface, we have:

\begin{equation}
\label{expandM2}
M^{OS} = \int_0^{2 \pi} \int_0^{\rho_{max}} \rho M^{OS}(\rho,\theta) d\rho
   d\phi =  
   M^{CS} + \Delta M + \delta M \,
\end{equation}
where $\rho_{max}$ is the radius of the M1 mirror. $\Delta M$
can be calculated numerically as the integral of $\Delta M(\rho,\theta)$:
\begin{eqnarray}
\label{DeltaM}
\Delta M(\rho,\theta)=-\alpha \left ( \begin{array}{ccc}
  \hat d_{1,p} \cos^2 \phi + \hat d_{1,s} \sin^2 \phi & ~~~~ &  (\hat d_{1,p}+\hat d_{1,s})
  \sin \phi \cos \phi \\ 
  (\hat d_{1,p}+\hat d_{1,s}) \sin \phi \cos \phi    &    & \hat d_{1,p} \sin^2 \phi +
  \hat d_{1,s} \cos^2 \phi
				\end{array} \right ) \nonumber \\
  + {\alpha^2 \over 2} \left ( \begin{array}{ccc}
  \hat d_{2,p} \cos^2 \phi + \hat d_{2,s} \sin^2 \phi & ~~~~ &  (\hat d_{2,p}+\hat d_{2,s})
  \sin \phi \cos \phi \\ 
  (\hat d_{2,p}+\hat d_{2,s}) \sin \phi \cos \phi    &    & \hat d_{2,p} \sin^2 \phi +
  \hat d_{2,s} \cos^2 \phi
				\end{array} \right ) + \dots
\, .
\end{eqnarray}

Mirror imperfections are accounted for by the (measured)
$M^{CS}$, whereas non-collimated incidence is accounted for by the
(calculated) $\Delta M$. $\delta M$ is an unknown second-order term that
couples mirror imperfections and non-collimated
incidence. This term is small (as shown below) and may be neglected for our
purposes here.  

The number of terms to retain in the Taylor expansion of Eq~(\ref{DeltaM})
depends on the particular 
telescope configuration and the accuracy required. Typical examples are
presented below in which $\Delta M$ can be neglected entirely (on-axis
mirror) or needs to be calculated up to second order (off-axis mirror, see
\S\ref{sec:offaxis}).

In the reminder of this section I present the results of numerical
simulations that provide some insight into the various terms that are
involved in the calibration procedure. The parameters of the simulation are
listed in Table~\ref{table:onaxis}. They represent a 4-m on-axis telescope
with a silver coating on the M1 mirror. The coating has been degraded 
%in two
%different ways. First, 
so that the complex refraction index fluctuates over the
mirror surface as $N(\rho,\phi)=N^{av} [1 + N^{f} cos (\phi/2)]$
($N^{av}$ is the 
average refraction index and $N^{f}$ is the amplitude of the
fluctuation). This choice has been made to represent a pessimistic 
scenario that induces a considerable amount of instrumental
polarization. The discretization of the simulation considers 100~points in
$\rho$ and 200 in $\phi$.
%The second difficulty introduced is a 10\% difference between
%the nominal refraction index $N^{nom}$ used for the theoretical calculation
%of $\Delta M$ and the actual average index $N^{av}$ of the mirror. In this
%manner the simulation incorporates our uncertainty on the refraction index
%of the mirror. 

The Jones matrices $M^{OS}$ and $M^{CS}$, obtained by applying
Eq~(\ref{explicit}) to each area element of the mirror, are:

\begin{equation}
\label{onMOS}
%M^{OS}= -0.97\exp(0.57i) \left [ {\bf 1} + \left ( \begin{array}{ccc} 
%        0.00 & ~~ & -0.01\exp(3.09i) \\
%        -0.01\exp(3.09i) &  & -3.16\times 10^{-4}\exp(0.20i) \end{array} \right ) 
M^{OS}= -0.94\exp(0.64i) \left [ {\bf 1} + \left ( \begin{array}{ccc} 
        0.00 & ~~ & -0.06\exp(1.72i) \\
        -0.06\exp(1.72i) &  & 4.20\times 10^{-5}\exp(0.40i) \end{array} \right ) 
  \right ] \, ,
\end{equation}

\begin{equation}
\label{onMCS}
%M^{CS}= M^{OS} + 2.80\times 10^{-6}\exp(1.83i) 
%  \left ( \begin{array}{ccc} 
%        0.00 & ~~ & -0.05\exp(0.53i) \\
%        -0.05\exp(0.53i) &  & -3.16\times 10^{-4}\exp(0.20i) \end{array} \right ) 
M^{CS}= M^{OS} + 1.56\times 10^{-5}\exp(0.78i) 
  \left ( \begin{array}{ccc} 
        1.00 & ~~ & -0.07\exp(0.28i) \\
        -0.07\exp(0.28i) &  & -0.76\exp(0.67i) \end{array} \right ) 
   \, .
\end{equation}

The ``irregularities'' introduced in the refraction index of the mirror break
the symmetry and give rise to polarizing effects in $M^{OS}$, with
off-diagonal terms of approximately 6\% . Fortunately,
the calibration setup matrix $M^{CS}$ is an excellent approximation to
$M^{OS}$, with a maximum difference of $\sim$10$^{-5}$. It is then possible
to calibrate the telescope almost down to the 
$10^{-5}$ level without even having to correct for the inclined incidence
(i.e., neglecting $\Delta M$ in Eq~[\ref{expandM2}])

\subsection{Off-axis mirror}
\label{sec:offaxis}

An off-axis mirror can be represented by a larger ``equivalent'' on-axis
mirror with a variable refraction index, as depicted in
Fig~\ref{fig:offaxis}. The equations derived in \S\ref{sec:onaxis} above are
still valid in the $(\rho,\phi)$ reference frame. One simply needs to set the
reflectivity to zero outside the shaded area of the figure
($r>r_{max}$). This can be accomplished, e.g. by setting the refraction index
to 1. 

This type of mirrors is slightly more complicated to calibrate. Even a
perfectly coated mirror will produce instrumental polarization. A simulation
with the parameters listed in Table~\ref{table:offaxis} yields the following
matrices: 

\begin{equation}
\label{offMOS}
M^{OS}= -0.97\exp(0.59i) \left [ {\bf 1} + \left ( \begin{array}{ccc} 
        0.00 & ~~ & 0.00 \\
        0.00 &  & -3.51\times 10^{-2}\exp(1.61i) \end{array} \right ) 
  \right ] \, ,
\end{equation}

\begin{equation}
\label{offMCS}
M^{CS}= M^{OS} + 1.24\times 10^{-2}\exp(2.22i) 
  \left ( \begin{array}{ccc} 
        1.00 & ~~ & 0.00 \\
        0.00 &  & 0.95\exp(3.11i) \end{array} \right ) 
   \, .
\end{equation}

$M^{CS}$ is now significantly different from $M^{OS}$ and we need to
calculate the correction term $\Delta M$ (Eqs~[\ref{expandM2}]
and~[\ref{DeltaM}]). Let us consider for the moment that the refraction index
is perfectly known ($\delta d_{i,p}=\delta d_{i,s}=0$ in
Eq~[\ref{splitd}]). Calculating $\Delta M$ with Eq~(\ref{DeltaM}) up to
second order we have that:

\begin{equation}
\label{offMCSDM}
M^{CS}+\Delta M = M^{OS} + 2.22\times 10^{-4}\exp(2.65i)  \left (
        \begin{array}{ccc}  
        1.00 & ~~ & 0.00 \\
        0.00 &  & 0.43\exp(0.44i) \end{array} \right ) 
\, .
\end{equation}

Let us now turn to the more general case of an off-axis mirror with surface
irregularities for which we have only an imperfect knowledge of the average 
refracion index. Again, we use a refracion index with an angular dependence
$N(\rho,\phi)=N^{av} [1 + N^{f} cos (\phi/2)]$. Furthermore, we do not know
exactly the average refraction index of the mirror $N^{av}$, but only an
approximation $N^{nom}$. This approximate value will be used in the
calculation of $\hat d_{i,p}, \hat d_{i,s}$ and $\Delta M$
(Eqs~[\ref{splitd}] and~[\ref{DeltaM}]). I carried out several experiments
with different values of $N^{nom}$ and $N^f$ to determine the sensitivity of
the calibration to these parameters. Some of the calculations
include only the first-order dependence of $\Delta M$ on $\alpha$ (first term
in the right-hand side of Eq~[\ref{DeltaM}]). These are denoted by $\Delta
M^*$ (as opposed to $\Delta M$, which considers the second-order term). 
Table~\ref{table:offaxis2} lists the results of these experiments. The first
two rows give the highest polarizing term in $M^{OS}$ for each
simulation. The third to fifth rows show how the calibration error decreases
with successive levels of approximation. By comparing the second and fifth
rows, we can see that the calibration is able to reduce the instrumental
polarization by an amount between one and two orders of magnitude.

\section{Conclusions}
\label{sec:conclusions}

This paper introduces a new concept to calibrate telescopes for astronomical
polarimetry. The proposed method is particularly useful for modern
large-aperture telescopes, for which it is probably the only practical
procedure (at least for purely instrumental calibration). An accurate
absolute calibration will be crucial for the new weak-signal science that the
ATST will open. Existing night-time telescopes may also take advantage of
this calibration procedure.

The Jones (and Mueller) matrix of an on-axis mirror is almost unaffected by
the non-collimated incidence of the beam in the CS. For the particular
configuration considered in \S\ref{sec:onaxis}, the Jones matrix obtained
from the calibration is good to almost $10^{-5}$ with no need for any
additional 
correction (see Eq~[\ref{onMCS}]). This is in spite of the relatively large
mirror imperfections in the simulation, which generate polarizing terms of
the order of 6\%. 

An off-axis mirror suffers more instrumental polarization due to the
asymmetric configuration. A mirror that produces instrumental polarization of
a few percent can be calibrated to reach the $10^{-4}$ level (see
Table~\ref{table:offaxis2}). In addition to measuring the Jones matrix of the
calibration setup ($M^{CS}$), it is also necessary to calculate the
correction term $\Delta M$. This calculation is straightforward, though, and
$\Delta M$ does not need to be recalculated unless the mirror is recoated or
it degrades to a point where its {\it average} refraction index changes
significantly. Note that the $10^{-4}$ calibration accuracy includes some
uncertainty on the average refraction index of the mirror.

\appendix
\section*{Appendix A: Mueller formalism}
\label{sec:appendix}
\setcounter{equation}{0}
\renewcommand{\theequation}{A{\arabic{equation}}}

I have used in this paper the Jones matrix formalism, which deals
directly with the components of the electric field of the light
wave. Sometimes, however, the Mueller formalism is more adequate, especially
when dealing with partially polarized or non-monochromatic light. Many
researchers are more familiar with the Mueller matrices and the Stokes
parameters. For these reasons it is probably useful to provide the Mueller
equivalent\cite{LdI02} of the Jones matrices derived in this work.

Eq~(\ref{onMOS}) becomes:

\begin{equation}
M^{OS} = \left ( \begin{array}{cccc}
   0.89 & -3.42\times 10^{-5} & 1.65\times 10^{-2} & -2.19\times 10^{-6} \\
  -3.42\times 10^{-5} & 0.88 & 5.49\times 10^{-7} & 0.11 \\
  1.65\times 10^{-2} & 5.49\times 10^{-7} & 0.89 & -1.45\times 10^{-5} \\
  2.19\times 10^{-6} & -0.11 & 1.45\times 10^{-5} & 0.88 \\
		 \end{array} \right ) \, .
\end{equation}

Eq~(\ref{onMCS}) becomes:

\begin{equation}
M^{CS} = M^{OS} + 2.23\times 10^{-5} \times \left ( \begin{array}{cccc}
0.30  &  1.00 & -6.77\times 10^{-2} & 6.42\times 10^{-2} \\
1.00 & 0.30 & -1.80\times 10^{-2} & 5.45\times 10^{-2} \\
-6.77\times 10^{-2} & -1.80\times 10^{-2} & 0.30 & 0.45 \\
-6.42\times 10^{-2} & -5.45\times 10^{-2} & -0.45 & 0.30 \\
					   \end{array} \right ) \, .
\end{equation}

Eq~(\ref{offMOS}) becomes:

\begin{equation}
M^{OS} = \left ( \begin{array}{cccc}  
       0.94 & -1.93\times 10^{-3} & 2.29\times 10^{-7} &  0.00   \\
  -1.93\times 10^{-3}   &  0.94 & 0.00  & 0.00    \\
  2.29\times 10^{-7} & 0.00  &   0.94   & 3.29\times 10^{-2}   \\
  0.00 & 0.00  & -3.29\times 10^{-2}   &  0.94   \\
		 \end{array} \right ) \, .
\end{equation}

Eq~(\ref{offMCS}) becomes:

\begin{equation}
M^{CS} = M^{OS} + 2.34\times 10^{-2} \times \left ( \begin{array}{cccc}
  6.92\times 10^{-3}  &  5.85\times 10^{-2}  & 0.00 & 0.00 \\
    5.85\times 10^{-2} &  6.92\times 10^{-3} & 0.00 & 0.00 \\
   0.00 & 0.00 &   2.96\times 10^{-2}  &   -1.00 \\
   0.00 & 0.00 &  1.00  &  2.96\times 10^{-2} \\
					    \end{array} \right ) \, .
\end{equation}

Eq~(\ref{offMCSDM}) becomes:

\begin{equation}
M^{CS} + \Delta M = M^{OS} + 1.82\times 10^{-4} \times \left (
\begin{array}{cccc} 
    -0.97  &  -0.13 & -1.78\times 10^{-5} & 0.00 \\
    -0.13  &  -0.97 & 0.00 & 1.13\times 10^{-5}  \\
 -1.78\times 10^{-5} & 0.00  &   -1.00  &   0.72  \\
  0.00 & -1.13\times 10^{-5}  &  -0.72  &   -1.00  \\
							       \end{array}
\right ) \, .
\end{equation}

\section*{Acknowledgments}
This work utilizes data from the Advanced Technology Solar
Telescope (ATST) project, managed by the National Solar Observatory,
which is operated by AURA, Inc. under a cooperative agreement with the
National Science Foundation.

\email{navarro@ucar.edu}

%\bibliographystyle{./osajnl.bst}
%\bibliography{../bib/aamnem99,../bib/articulos}

\clearpage

\begin{table}
%  \tablewidth{0pt}
  \caption[]{Simulation of a 4-m on-axis mirror}
     \label{table:onaxis}
     $$
  \begin{array}{cc}
    \hline
%    \noalign{\smallskip}
  Parameter & Value \\
%  \noalign{\smallskip}
  \hline
%  \noalign{\smallskip}
  \rho_{max} & 2~m \\
   F   & 8~m \\
   H_{cal}   &  10~m \\
   N^{av}  & 0.2 + 3.4 i  \\
   N^{f}  & 0.50 \\
  max(   | M^{OS} - M^{CS} |   ) & 1.56\times 10^{-5} \\
 % $N^{nom}$ & 1.10 $N^{av}$ \\
  \hline
  \end{array}
  $$
\end{table}

\clearpage

\begin{table}
%  \tablewidth{0pt}
  \caption[]{Simulation of a 4-m off-axis mirror}
     \label{table:offaxis}
     $$
  \begin{array}{cc}
    \hline
%    \noalign{\smallskip}
  Parameter & Value \\
%  \noalign{\smallskip}
  \hline
%  \noalign{\smallskip}
   r_{max}  & 2~m \\
   \rho_{max}  & 6~m \\
   F   & 8~m \\
   H_{cal}   &  10~m \\
   N^{av}  & 0.2 + 3.4 i  \\
   N^{f}  & 0.0 \\
  max(   | M^{OS} - M^{CS} |   ) & 1.24\times 10^{-2} \\
  \hline
  \end{array}
  $$
\end{table}

\clearpage

\begin{table}
%  \tablewidth{0pt}
  \caption[]{Simulation of a 4-m off-axis mirror with surface irregularities}
     \label{table:offaxis2}
     $$
  \begin{array}{ccccc}
    \hline
%    \noalign{\smallskip}
                    & N^f=0.10 & N^f=0.10 & N^f=0.25 & N^f=0.25 \\
                    & N^{nom} = N^{av} &  N^{nom}=1.01 N^{av}  & 
    N^{nom}=1.01 N^{av} & N^{nom}=1.10 N^{av}  \\
  \hline
% $|M^{OS}_{2,2}/M^{OS}_{1,1}|-1$ &  &  &  &  \\
 |M^{OS}_{2,2}|-|M^{OS}_{1,1}| & 2.02\times 10^{-3} & 2.02\times
   10^{-3} & 2.25\times 10^{-3}   &   2.25\times 10^{-3}  \\
    |M^{OS}_{1,2}|               &  1.50\times 10^{-2}  &  1.50\times 10^{-2}  & 3.86\times 10^{-2}   &  3.86\times 10^{-2}   \\
   max(|M^{OS}-M^{CS}| )        &  1.25\times 10^{-2}  &  1.25\times 10^{-2}  & 1.26\times 10^{-2}   &  1.26\times 10^{-2} \\
   max[ |M^{OS}-  \\
    (M^{CS} + \Delta M^*)| ]     &  4.04\times 10^{-3}  &  3.89\times
   10^{-3}  &  3.71\times 10^{-3}  &  2.59\times 10^{-3}  \\ 
   max[ |M^{OS}-  \\
    (M^{CS} + \Delta M)| ]       &  2.84\times 10^{-4}  &  4.18\times 10^{-4}  &  7.11\times 10^{-4}  &  1.80\times 10^{-3} \\
  \hline
   \end{array}
  $$
\end{table}

\clearpage

\begin{figure}[t]
%\centerline{\includegraphics{fig1.eps}}
\centerline{}
\caption{\label{fig:setups}
Left: Normal observing configuration (OS), with a collimated incident
      beam. Right: Calibration configuration (OS), with an inclined
      (diverging) incident beam. navarrof1.eps.
}
\end{figure}

\begin{figure}[t]
%\centerline{\includegraphics{fig1.eps}}
\centerline{}
\caption{\label{fig:angles}
Incidence angles $\theta$ and $\alpha$ for the OS and CS,
respectively. Solid (dashed) represent rays in the OS (CS). The dotted line
represents the normal to the mirror surface. navarrof2.eps.
}
\end{figure}

\begin{figure}[t]
%\centerline{\includegraphics{fig1.eps}}
\centerline{}
\caption{\label{fig:offaxis}
Schematic representation of the equivalent on-axis mirror. Left: Lateral
view. The thick line represents the actual off-axis mirror of radius
$r_{max}$. The thin line represents the equivalent on-axis mirror. Right:
Top-down view. The shaded area is the actual off-axis mirror. navarrof3.eps.
}
\end{figure}

\clearpage

\begin{figure*}
\includegraphics[width=0.8\textwidth]{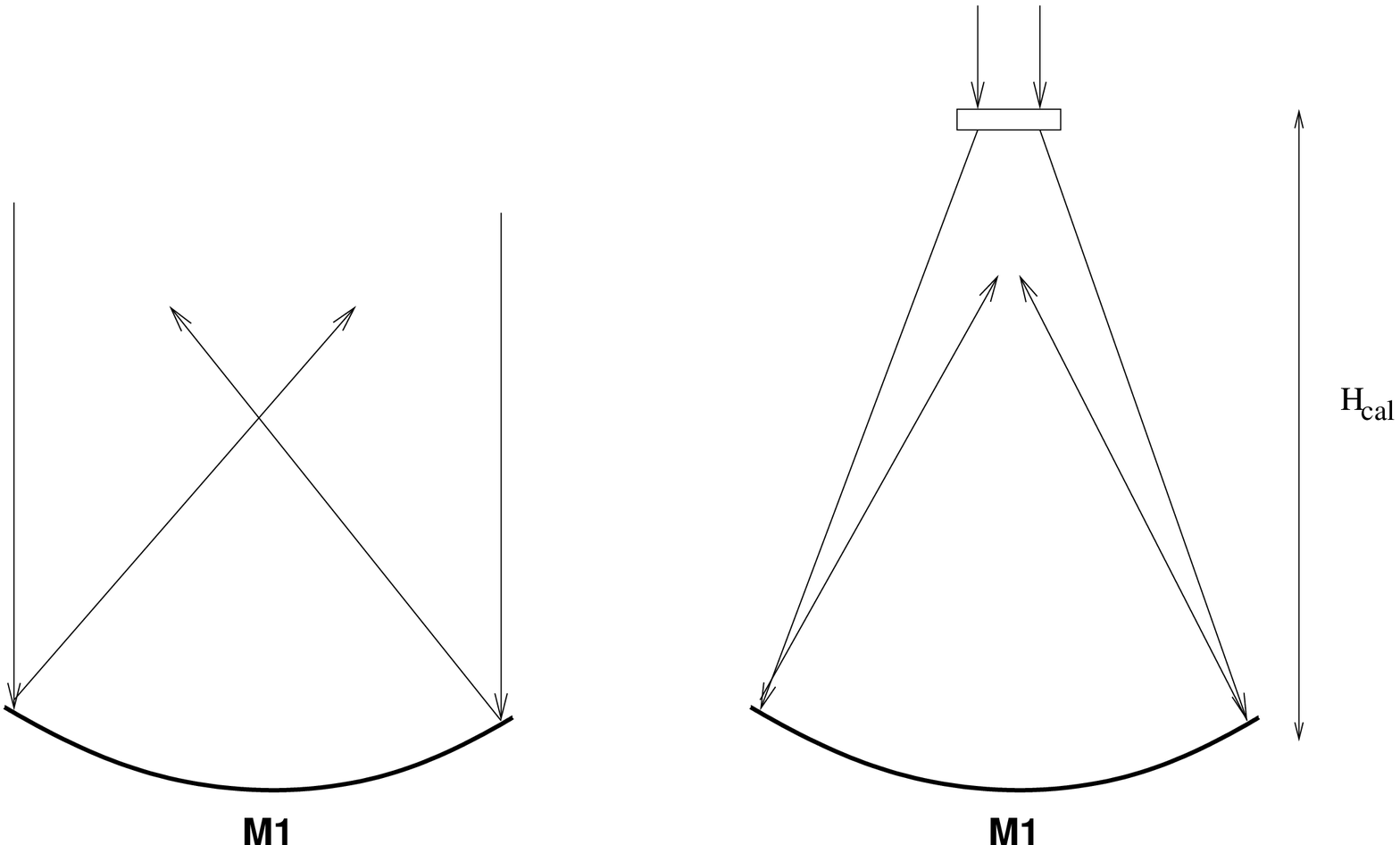}
%\caption{
%\label{fig:setups}
%Left: Normal observing configuration (OS), with a collimated incident
%      beam. Right: Calibration configuration (OS), with an inclined
%      (diverging) incident beam.}
\end{figure*}

\clearpage

\begin{figure}
\includegraphics[width=0.8\textwidth]{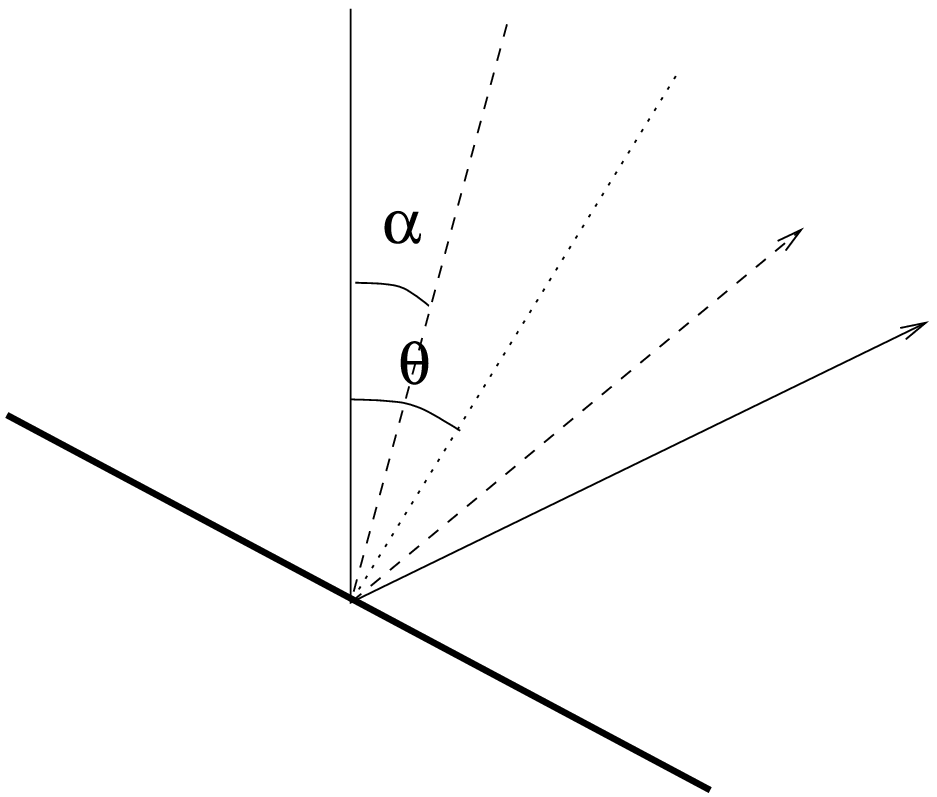}
%\caption{
%\label{fig:angles}
%Incidence angles $\theta$ and $\alpha$ for the OS and CS,
%respectively. Solid (dashed) represent rays in the OS (CS). The dotted line
%represents the normal to the mirror surface.
%}
\end{figure}

\clearpage

\begin{figure*}
\includegraphics[width=0.8\textwidth]{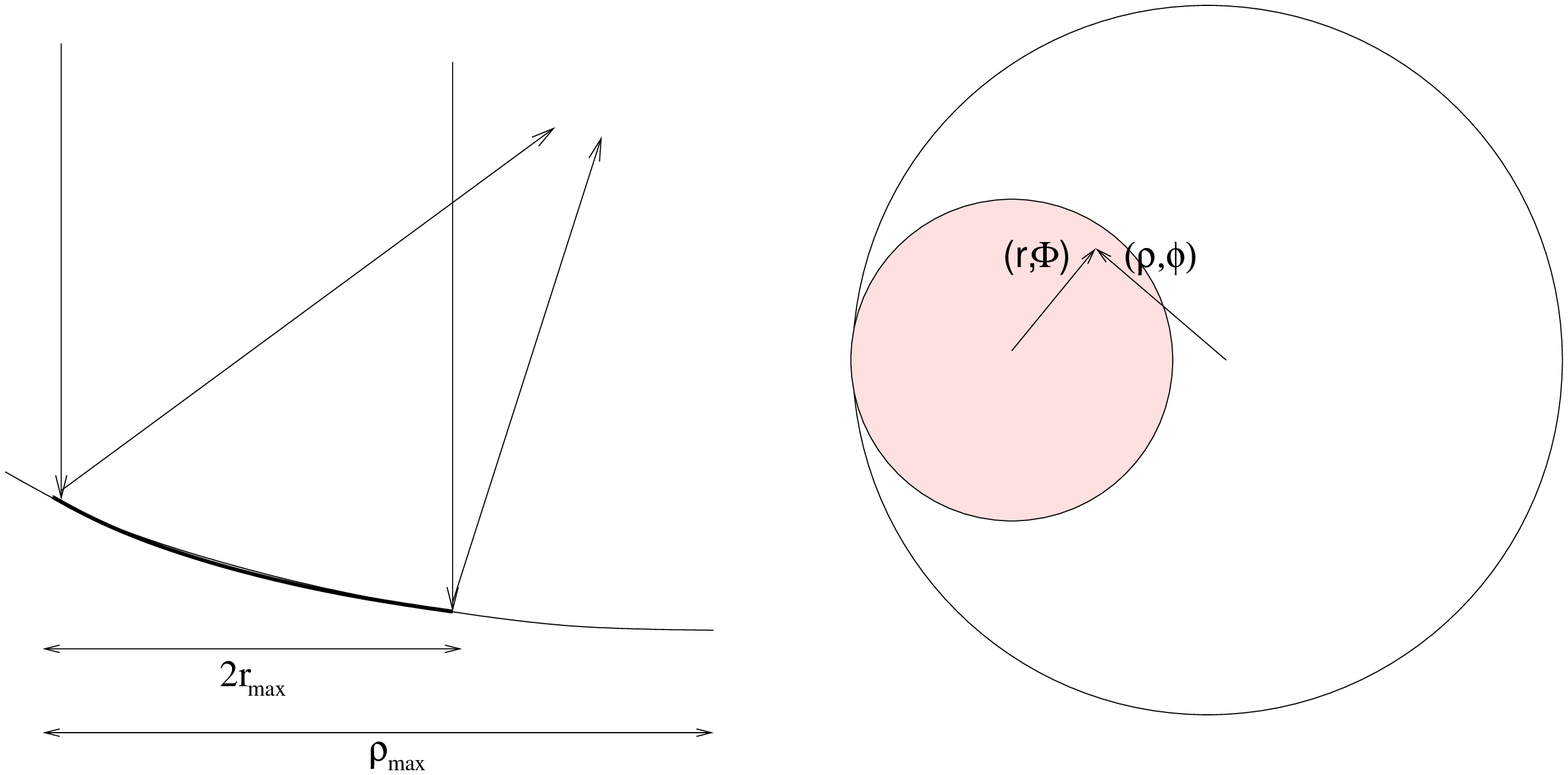}
%\caption{
%\label{fig:offaxis}
%Schematic representation of the equivalent on-axis mirror. Left: Lateral
%view. The thick line represents the actual off-axis mirror of radius
%$r_{max}$. The thin line represents the equivalent on-axis mirror. Right:
%Top-down view. The shaded area is the actual off-axis mirror.
%}
\end{figure*}

\clearpage

%\appendix

%\section*{Appendix A: Sample}
%\setcounter{equation}{0}
%\renewcommand{\theequation}{A{\arabic{equation}}}

%\begin{equation}
%a+b=c.
%\end{equation}

%\begin{thebibliography}{99}
%%Do not include separate BibTeX files; if BibTeX is used,
%% paste the output (contents of .bbl file) here.

%\bibitem{1} ...

%\bibitem{2} ...

%\end{thebibliography}

\end{document}